\documentclass[conference]{IEEEtran}
\IEEEoverridecommandlockouts
\usepackage{cite}
\usepackage{amsmath,amssymb,amsfonts}
\usepackage{algorithmic}
\usepackage{graphicx}
\usepackage{textcomp}
\usepackage{xcolor}
\usepackage[draft]{changes}
\definecolor{grape}{rgb}{0.63, 0.15, 0.75}
\definechangesauthor[color=grape]{ee}

\usepackage{enumitem}
\def\BibTeX{{\rm B\kern-.05em{\sc i\kern-.025em b}\kern-.08em
    T\kern-.1667em\lower.7ex\hbox{E}\kern-.125emX}}

\begin{document}

\title{Enabling Micro-payments on IoT Devices using Bitcoin Lightning Network}

\author{\IEEEauthorblockN{
Ahmet Kurt\IEEEauthorrefmark{1},
Suat Mercan\IEEEauthorrefmark{1},
Enes Erdin\IEEEauthorrefmark{2}, and
Kemal Akkaya\IEEEauthorrefmark{1}}
\IEEEauthorblockA{\IEEEauthorrefmark{1}Dept. of Electrical and Computer Engineering, Florida International University, Miami, Florida\\
Email: \{akurt005, smercan, kakkaya\}@fiu.edu}
\IEEEauthorblockA{\IEEEauthorrefmark{2}Dept. of Computer Science, University of Central Arkansas, Conway, Arkansas\\
Email: eerdin@uca.edu}}

\IEEEoverridecommandlockouts
\IEEEpubid{\makebox[\columnwidth]{978-0-7381-1420-0/21/\$31.00~\copyright2021 IEEE \hfill} \hspace{\columnsep}\makebox[\columnwidth]{ }}

\maketitle


\begin{abstract}
Lightning Network (LN) addresses the scalability problem of Bitcoin by leveraging off-chain transactions. Nevertheless, it is not possible to run LN on resource-constrained IoT devices due to its storage, memory, and processing requirements. Therefore, in this paper, we propose an efficient and secure protocol that enables an IoT device to use LN's functions through a gateway LN node. The idea is to involve the IoT device in LN operations with its digital signature by replacing original 2-of-2 multisignature channels with 3-of-3 multisignature channels. Our protocol enforces the LN gateway to request the IoT device's cryptographic signature for all operations on the channel. We evaluated the proposed protocol by implementing it on a Raspberry Pi for a toll payment scenario and demonstrated its feasibility and security. 
\end{abstract}


\section{Introduction}
Internet of Things (IoT) has been adopted in various domains at a great pace in the last decade as it brings numerous opportunities and convenience \cite{gubbi2013internet}. We have been witnessing applications where an IoT device may need to do financial transactions. For instance, they may be used in commerce applications such as toll systems, where an on-board unit acting as an IoT device on a vehicle may need to do automatic payments as the vehicle passes through a toll gate \cite{pavsalic2016vehicle}. Similarly, there are other cases such as automated vehicle charging, parking payment, sensor data selling, etc. where \textit{micro-payments} need to be made \cite{mercan2021cryptocurrency, robert2020enhanced, kurt2020lnbot}. In this context, cryptocurrencies have great potential to provide a smooth payment automation. Thus, successful merge of IoT \cite{mercan2020improving} and cryptocurrency technologies such as Bitcoin \cite{nakamoto2019bitcoin} and Ethereum \cite{wood2014ethereum} may foster such applications.

However, despite their popularity, mainstream cryptocurrencies such as Bitcoin and Ethereum suffer from scalability issues \cite{xie2019survey} in terms of transaction confirmation times and throughput \cite{throughput}. This increases the transaction fee and makes their adoption infeasible for micro-payments. Payment Channel Network (PCN) idea has emerged as a second layer solution to address this problem by utilizing off-chain transactions \cite{decker2015fast}. While LN is a successful solution that addresses the scalability issue, current LN protocol cannot be run on most of the IoT devices because of the computation, communication, and storage requirements \cite{lniotproblem}. 

To this end, in this paper, we propose an efficient and secure protocol where an IoT device can connect to an \textit{LN gateway} that already hosts the full LN and Bitcoin nodes and can send payments on behalf of the IoT device when requested. First, a secure channel is established between IoT device and the LN gateway using lightweight cryptography. Second, we introduce the concept of \textit{3-of-3 multisignature LN channels}, which involve signatures of all parties (i.e., the IoT device, the LN gateway and a bridge LN node to which the gateway opens a channel) to conduct any operation on the channel as opposed to using LN's original 2-of-2 multisignature channels \cite{2of2multisig}. More specifically, the LN gateway cannot spend channel funds without getting IoT device's cryptographic signature. 

In order to assess the effectiveness and overhead of the proposed protocol, we implemented it within a setup where a Raspberry Pi sends payments to a real LN node through a wireless connection. A toll payment scenario was considered for the evaluation. We demonstrated that the proposed protocol enables realization of timely payments without compromising the security of IoT device's funds.

\section{System Model and Protocol}
\label{sec:systemmodel}

\subsection{System Model}

\begin{figure}[h]
    \centering
    \vspace{-2mm}
    \includegraphics[width=\linewidth]{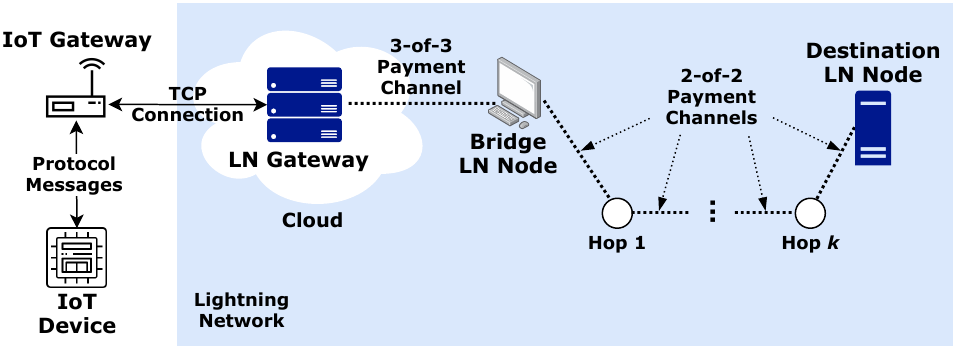}
    \vspace{-6mm}
    \caption{\small Illustration of our assumed system model.}
    \label{fig:system_model}
    \vspace{-6mm}
\end{figure}

There are five entities in our system which are \textbf{IoT device}, \textbf{IoT gateway}, \textbf{LN gateway}, \textbf{bridge LN node}, and \textbf{destination LN node} as shown in Fig. \ref{fig:system_model}. IoT device wants to pay the destination LN node for the goods/services. The IoT gateway is responsible for connecting the IoT device to the Internet when the IoT device is within its range using wireless protocols such as WiFi, Bluetooth, 5G etc. Through this Internet connection, IoT device is able to reach the LN gateway, which manages the Bitcoin and LN nodes and can be hosted anywhere on the cloud. Bridge LN node is the node to which the LN gateway opens a channel when requested by the IoT device. This node could be any LN node on the Internet and is determined by the LN gateway. Through the bridge LN node, IoT device's payments are routed to a destination LN node specified by the IoT device. 

\subsection{Proposed Protocol}
\subsubsection{Channel Opening Process}

We authorize the IoT device to securely initiate the channel opening process through the LN gateway and co-sign the funding transaction along with the LN gateway and bridge LN node. This means that LN's existing channel opening protocol is modified with the addition of another party (i.e., IoT device) to the channel. Since now three signatures are required to open a channel, the channel is no longer a 2-of-2  multisignature wallet but instead a \textit{3-of-3 multisignature} wallet. To create a secure channel between IoT and the LN gateway, we use public key cryptography and symmetric session keys to exchange messages. IoT sends a channel opening request message to the LN gateway. The LN gateway creates a Bitcoin transaction from the IoT device's BTC wallet address to the 3-of-3 multisignature address. The LN gateway also creates the commitment transactions for itself and the bridge LN node. The LN gateway then sends the funding transaction to IoT to get it completely signed. Finally, the LN gateway broadcasts the signed funding transaction to the Bitcoin network which opens the channel.

\subsubsection{Sending a Payment}

We incorporate the IoT device in the process for signature aggregation. In addition, the gateway needs to charge the IoT a \textit{service fee} for its services. To start the payment sending, IoT device sends a payment sending request to the LN gateway. Upon receiving the request, the LN gateway adds an HTLC output to its commitment transaction. When preparing the HTLC, the LN gateway \textit{deducts a certain amount of fee} from the real payment amount IoT device wants to send to the destination. Therefore, the remaining Bitcoin is sent with the HTLC. The new commitment transaction created at the LN gateway needs to be also signed by IoT device and thus we enforce the gateway to make a request to IoT device. Upon receiving this request, IoT device signs the commitment transaction and sends it back to the LN gateway. Next, the LN gateway and bridge LN node cross signs their new commitment transactions and revoke the old state. This finalizes the payment sending and IoT is notified of the successful payment.

\subsubsection{Channel Closing Process}

All 3 parties of the channel namely; The IoT device, the LN gateway and the bridge LN node can close the channel. The IoT device can ask the LN gateway to close the channel by sending a channel closing request. The LN gateway can also close the channel by requesting from IoT device to sign its most recent commitment transaction then broadcasting the signed transaction to the blockchain.

\section{Evaluation}
\label{sec:evaluation}

\subsection{Experiment Setup}

To evaluate the proposed protocol, we created a setup where an IoT device connects to an LN gateway to send payments on LN. To mimic the IoT device, we used a Raspberry Pi 3 Model B v1.2 and the LN gateway was setup on a desktop computer with an Intel(R) Xeon(R) CPU E5-2630 v4 and 32 GB of RAM. This desktop computer was on a remote location different from the Pi. For the full Bitcoin node installation, we used \textit{bitcoind} \cite{bitcoind} and for the LN node, we used \textit{lnd v0.11.0-beta.rc1} from Lightning Labs \cite{lnd} which is a complete implementation of the LN protocol. Python was used to implement the protocol. We used IEEE 802.11n (WiFi) to exchange protocol messages between the Raspberry Pi and the LN gateway.

\subsection{Toll Payment Use-Case Evaluation of the Protocol}

When a car enters the communication range of the toll wireless system, it sends a request to the IoT gateway of the toll to initiate the payment. IoT gateway relays this request to the LN gateway which immediately creates an LN payment and sends it to the toll center's LN node. Lastly, the car is notified of the successful payment. For this to work, whole payment sending process must be completed before the car gets out of the communication range of toll's wireless system.

There are 4 protocol message exchanges between IoT device and the LN gateway in our payment sending protocol. In each protocol message exchange, there are 2 corresponding communication delays which are between the car and the IoT gateway and between the IoT gateway and the LN gateway running at the cloud. In addition to communication delays, we also measured the LN payment sending time which was 2 seconds over an average of 30 separate payments at different times throughout the day. Eventually, the \textit{total payment sending time for WiFi was 2.558 secs}.

\begin{table}[h]
 \vspace{-3mm}
  \begin{center}
    \caption{Available Time for Different Vehicle Speeds}
     \vspace{-2mm}
    \label{tab:toll}
    \resizebox{0.65\linewidth}{!}{
    \begin{tabular}{|c|c|c|}
      \hline
     \textbf{Vehicle Speed}  &  \textbf{Available Time} & \textbf{Satisfied?}  \\
     \hline
      50 mph  &  11.2 s   &  Yes \\ 
      \hline     
      60 mph  &  9.3 s   &  Yes  \\ 
      \hline
      80 mph  &  7 s     &  Yes  \\
     \hline
    \end{tabular}
  }
  \vspace{-3mm}
  \end{center}
\end{table}

The advertised range of 802.11n is 250m \cite{802.11nrange}. If cars pass through the toll gate with a speed of 50mph, there is around 11 seconds with WiFi available for them to complete the protocol message exchanges with the LN gateway for a successful toll payment. The results for varying vehicle speeds are shown in Table \ref{tab:toll}. As can be seen, our protocol meets the deadlines even under a high speed of 80mph.

\section{Conclusion}
\label{sec:conclusion}
In this paper, we proposed a secure and efficient protocol for enabling IoT devices to use Bitcoin's LN for sending payments. By modifying LN's peer protocol for channel management as well as LN's Bitcoin transactions, a third peer (i.e. IoT device) was added to LN channels. IoT device's interactions with LN is achieved through a gateway node which has access to LN and thus can provide LN services to it in return for a fee. Our evaluation results showed that the proposed protocol enables smooth payments for the IoT device with negligible delay/cost overheads while ensuring the security of the IoT device's funds.

\bibliography{references}
\bibliographystyle{IEEEtran}

\end{document}